\documentclass[aps,prl,twocolumn,preprintnumbers,showpacs,superscriptaddress,nobalancelastpage]{revtex4}

\begin{document}

\title{Testing the Collective Properties of Small-World Networks
through Roughness Scaling}

\author{B. Kozma}
\email{kozmab@rpi.edu}
\affiliation{Department of Physics, Applied
Physics, and Astronomy, Rensselaer Polytechnic Institute, 110
8$^{th}$ Street, Troy, NY 12180--3590, USA}

\author{M. B. Hastings}
\email{hastings@cnls.lanl.gov}
\affiliation{Center for Non-linear Studies and Theoretical Division,
Los Alamos National Laboratory, Los Alamos, NM 87545}

\author{G. Korniss}
\email{korniss@rpi.edu}
\affiliation{Department of Physics, Applied
Physics, and Astronomy, Rensselaer Polytechnic Institute, 110
8$^{th}$ Street, Troy, NY 12180--3590, USA}

\begin{abstract}
Motivated by a fundamental synchronization problem in
scalable parallel computing and by a recent criterion for
``mean-field'' synchronizability in interacting systems, we study
the Edwards-Wilkinson model on two variations of a small-world
network. In the first version each site has exactly one random link of
strength $p$, while in the second one each site on average has $p$
links of unit strength. We construct a perturbative description for the width
of the stationary-state surface (a measure of synchronization),
in the weak- and sparse-coupling limits, respectively,
and verify the results by performing exact
numerical diagonalization. The width remains finite in both cases, but
exhibits anomalous scaling with $p$ in the latter for $d\leq 2$.
\end{abstract}


\date{\today}
\maketitle

Crossovers from low-dimensional to mean-field-like behavior have
been studied and found in various interacting systems
\cite{BARRAT,GITTERMAN,xy_sw,ising_sw,phase_sw} when the
(original) regular and short-range interaction topology is
modified to a small-world (SW) network
\cite{WATTS98,SW_book,NEWMAN}. Mean-field behavior is commonly
observed, even when the random links are added to a
one-dimensional original ``substrate''
\cite{xy_sw,ising_sw,phase_sw}. Recently, a general criterion for
the crossover to mean-field behavior was given by Hastings
\cite{HASTINGS03}.

The equilibrium Edwards-Wilkinson (EW) model \cite{EW}
(considering it as the $\phi^2$ Gaussian model at the continuous
phase transition) is a particularly interesting case in that it
does {\em not} satisfy the mean-field criterion \cite{HASTINGS03}
when each pair of sites is connected with probability $p/N$ with a
link of unit strength on top of a {\em one-dimensional} regular
substrate. Thus, the EW model on this SW network is expected to
display drastically different scaling properties (as a function of
$p$) than the mean-field version of the model in one dimension. In
this paper we address this question.

The scaling properties of the EW model on regular or random
networks have direct relevance to the scalability and
synchronizability of parallel discrete-event simulations
(PDES)\cite{KORNISS00,KORNISS03a}. In PDES schemes the individual
processing elements (PE) generate their own time streams for
update attempts (local simulated times) \cite{FUJI,LUBA}. The
synchronization/communication between neighboring PEs (following
the interaction topology of an underlying short-range interacting
system with asynchronous dynamics) leads to Kardar-Parisi-Zhang
(KPZ)-like kinetic roughening \cite{KPZ,BARABASI} of the
simulated time horizon \cite{KORNISS00,KORNISS_MRS}. For a
one-dimensional chain or ring of PEs, the steady-state performance
of the PDES scheme is governed by the EW Hamiltonian
\cite{KORNISS00}. In particular, for $N$ PEs, the width of the
simulated time horizon {\em diverges} as $N^{1/2}$, seriously
hindering efficient data collection and state saving
\cite{KORNISS03a,KORNISS_ACM}. Since scalable data management
crucially depends on the finiteness of the width of the time
horizon (the spread of the progress of the individual PEs), one
must suppress the diverging fluctuations of the simulated time
horizon. As an alternative to costly and frequent global
synchronizations among the PEs, an autonomous small-world
synchronization scheme was demonstrated to work \cite{KORNISS03a}.
This finding provided another concrete example for
synchronizability in generalized multi-agent systems facilitated
by a small-world network \cite{WATTS98,Strogatz_review}.

Motivated by both the generic scaling properties of SW networks,
and also the specific applications to scalable PDES
synchronization schemes, we will consider two different variations
of a the EW model on a SW network. In the ``soft'' and frequently
studied version of the SW network, random links of unit
strength are added to the one-dimensional substrate with
probability $p/N$ to each pair of sites
\cite{NEWMAN,NEWMAN_WATTS,MONA}. In the ``hard'' version, each site
has {\em exactly one} random link (in addition to the
nearest-neighbors) and the strength of the interaction through the
random links is $p$. That is, pairs of sites are selected at
random, and once they are linked, they cannot be selected again.
The latter construction originates from scalable PDES schemes,
where fluctuations in the individual connectivity of the PEs are
to be avoided \cite{KORNISS03a}. We introduced the above
terminology for the two versions of the network in regard to the
eigenvalue spectrum of the respective coupling matrix \cite{MONA},
discussed at the end.

We consider the equation (for a single realization of the small world)
\begin{equation}
\partial_{t} h_i =
-(2h_i-h_{i+1}-h_{i-1}) - \sum_{j=1}^{N} J_{ij}(h_i-h_j) +
\eta_{i}(t)\;, \label{EW_sw}
\end{equation}
where $h_i$ is the surface height, $\eta_{i}(t)$ is a
delta-correlated Gaussian noise with variance $2$ (without loss of
generality), and we have dropped the $t$-dependence from the
argument of $h_i$ for brevity. The {\em symmetric} matrix $J_{ij}$
represents the quenched random links on top of a one-dimensional
lattice of length $N$ with periodic boundary conditions. For the
hard version of the small world $J_{ij}$ has exactly one non-zero
element (being equal to $p$) in each row and column. This
construction results in $\sum_{l}J_{il}=p$ for {\em all} $i$. For
the soft version of the small world, each element of $J_{ij}$
(e.g., above the diagonal) is 1 with probability $p/N$ and zero
otherwise. In this case $[\sum_{l}J_{il}]=p$, where $[\ldots]$
denotes the average over the network disorder, i.e., the {\em
average} coordination number is $p$. We write Eq.~(\ref{EW_sw}) as
$\partial_t h_i=-\sum_{j}\Gamma_{ij} h_j + \eta_i$,
where $\Gamma=\Gamma^o+V$. Here, $\Gamma^o$ is the Laplacian of
the original one-dimensional ring, while
$V_{ij}=-J_{ij}+\delta_{ij}\sum_{l}J_{il}$ is the Laplacian on the
random part of the network.

In this Letter we focus on the behavior of the width, which
probes the generic collective properties of the underlying networks by
providing a sensitive measure of synchronization \cite{KORNISS03a}.
For a given realization of the small-world network the average
surface width characterizing the roughness is equal to
\begin{equation}
\langle w^2 \rangle_{N} \equiv
\left\langle\frac{1}{N}\sum_{i=1}^{N}(h_i-\bar{h})^2\right\rangle
= \frac{1}{N}\sum_{k=1}^{N-1}\frac{1}{\lambda_k} \;,
\label{w2}
\end{equation}
where $\bar{h}$$=$$(1/N)\sum_{i=1}^{N}h_i$ is the mean height,
$\langle\ldots\rangle$ denotes an ensemble average over the noise
in Eq.~(\ref{EW_sw}), and $\lambda_k$ are the non-zero eigenvalues
of the real symmetric coupling matrix $\Gamma$.
Our numerical
scheme relied on the exact diagonalization \cite{NUM_REC} of the
coupling matrix $\Gamma$ for a given realization of the network,
and then we exploited the right side of Eq.~(\ref{w2}) to obtain
the width. Finally, we obtained the {\em disorder-averaged}
(denoted by $[\ldots]$) width $[\langle w^2 \rangle_{N}]$ by
averaging over a large number of (ranging from $100$ to
$1000$) realizations.

\begin{figure}[t]
\vspace*{2.0truecm}
       \includegraphics{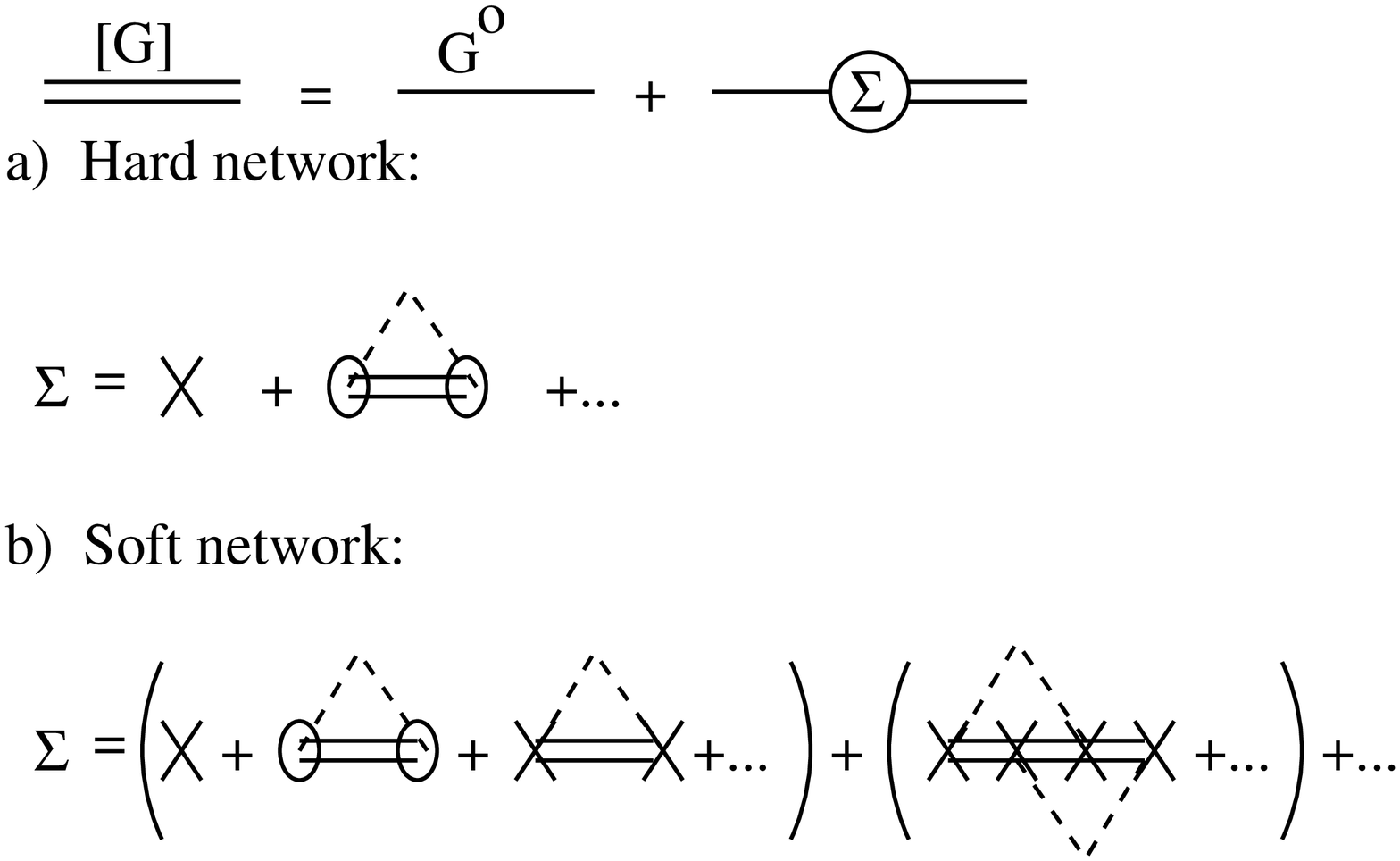}
\vspace*{3.0truecm}
\caption{Diagrams for the calculation of the
Green's function and the self-energy.}
\label{fig1}
\end{figure}

To motivate the results that follow, we consider the role of
dimension $d$ for the EW model on the soft version of the network,
along the lines of Ref.~\cite{HASTINGS03}.  As an approximation to
the soft network, consider instead an alternate model, where
in addition to the
nearest-neighbor connections, each site is coupled to {\em all} others
with strength $p/N$.
This yields the trivial (inhomogeneous
Landau) mean-field behavior with the mass $p$ \cite{meaf} and the
corresponding correlation length $\xi \sim p^{-1/2}$, and a
correlation volume $\xi^d\sim p^{-d/2}$.
Now, let us check if this model is a valid description
of the soft network.
For the soft network,
within this volume one would have on average $p \xi^d \sim p^{1-d/2}$ links.
For $d$$>$$2$ this number diverges as $p$$\rightarrow$$0$; then for
small $p$ there are a large number of links leaving the volume and
the sample-to-sample fluctuation in the mass in a correlation
volume is negligible compared to the mass itself so that the
trivial mean-field behavior is expected to be valid.
For $d$$\leq$$2$, the trivial behavior breaks down as we will see in
a perturbative calculation. In contrast, for the hard version of
the network the density of random links is unity, and the
trivial mean-field scaling is expected to hold for all $d$ in the
$p$$\rightarrow$$0$ limit.

The same result can be obtained by using the general criterion \cite{HASTINGS03}
in terms of the critical exponents for the correlation length and for the
susceptibility, $\nu$ and $\gamma$, respectively.
When viewed as a model near a continuous phase transition, the EW model
has $\nu$$=$$1/2$, $\gamma$$=$$1$.  The criterion $\nu d/2+\gamma/2>1$
for the validity of mean-field behavior is thus violated for $d$$=$$1$,
while $d$$=$$2$ is marginal.

We define the propagator $G$ to be equal to $\Gamma^{-1}$ in the space
of non-zero eigenvalues of $\Gamma$, while $G$ vanishes when acting
on the zero mode of $\Gamma$.  Thus, $G=P(\Gamma+i\epsilon)^{-1}$, where
$P_{ij}=\delta_{ij}-1/N$ is the projector onto the vector space orthogonal
to this zero mode.
We now present the results of a perturbative approach for the disorder-averaged
propagator $[G]$, using techniques of impurity
averaged perturbation theory \cite{iapt}, leaving details to be
published elsewhere \cite{future}.
The perturbative
expansion of $[G]$ can be obtained by
$[G]=G^o-[G^o V G]=G^o-[G^o V G^o] + [G^o V G^o V G^o] -...$,
where $G^o$$=$$P(\Gamma^o+i\epsilon)^{-1}$ is the
propagator of the Laplacian on the original one-dimensional
lattice. To obtain $[G]$ it is necessary to average this expansion
over the network disorder in $V$.

To deal with only
one-particle irreducible disorder-averaged diagrams, we calculate
the self-energy $\Sigma = ([G])^{-1} - (G^o)^{-1}$ perturbatively
for both the soft and hard versions of the small-world.
In Fig.~\ref{fig1}, a single line denotes the
propagator $G^o$ while a double line denotes $[G]$.
The relation between $[G],G^o$, and $\Sigma$ is shown at the top of
the figure.
In these calculations, a cross with no dashed lines attached is used to
denote the average over different realizations of the network of a diagonal term
$V_{ii}$ in $V$.  A pair of crosses connected by
a dashed line is used to denote an average
$[V_{ii}V_{jj}]-
[V_{ii}]
[V_{jj}]$, while three or more crosses connected by
dashed lines are used to denote higher cumulants.  Similarly,
circles connected by dashed lines are
used to denote averages of off-diagonal terms
$-J_{ij}$ in $V$.  Dashed lines can connect both circles and crosses.
Terms in the expansion with a circle, not connected by
dashed lines to other circles, vanish as $1/N$ for large $N$ and so may be
neglected.  For the hard network, terms in the expansion with two or more
crosses connected by dashed lines vanish since there are no random fluctuations
in $\delta_{ij}\sum_{l}J_{il}$ in this case, while for the soft network
these terms do appear.  In the hard network, each power of $V$ comes
with a power of $p$, while in soft network, each set of circles or crosses
connected by dashed lines comes with a power of $p$.

\begin{figure}[t]
\vspace*{2.5truecm}
       \includegraphics{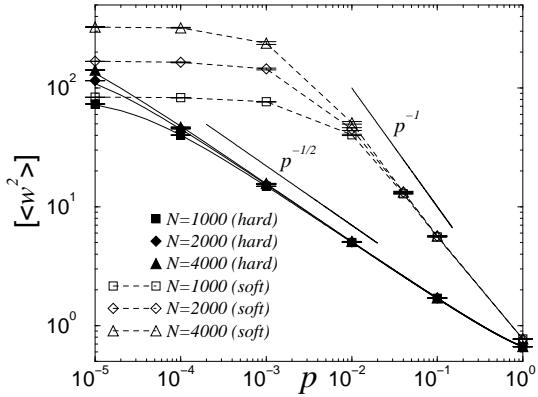}
\vspace*{3.0truecm}
\caption{Disorder-averaged width obtained by exact numerical
diagonalization for the hard (filled symbols) and for the soft (open symbols
with dashed lines to guide the eye)
version of the small world as a function $p$ for system sizes indicated in the
figure. The two slopes indicate the asymptotic small-$p$ infinite system-size
behavior. The solid lines are obtained using the finite-system propagator
with the effective mass Eq.~(\ref{sigma_hard}) for the hard network.}
\label{fig2}
\end{figure}

In the hard version of the network, we can proceed by expanding
$[G]$ in powers of $V$. The first diagram in
Fig.~\ref{fig1}(a) yields the lowest order result $\Sigma=p$.
Adding the second diagram yields
$\Sigma=p-p^2 [G]_{ii}=p-p^2/(2\sqrt{\Sigma})$,
where we use the fact that, for large $N$, $[G]_{ii}=1/(2\sqrt{\Sigma})$
plus terms of order $\sqrt{\Sigma}$ (see, e.g., \cite{TORO}).
Inserting the leading order result for $\Sigma$ into the second diagram
we find
\begin{equation}
\Sigma_{\rm hard} = p-(1/2)\, p^{3/2} + \ldots\;,
\label{sigma_hard}
\end{equation}
so that the higher powers of $V$ lead to corrections to $\Sigma$ which
are higher order in $p$ as required.

In the soft version of the network, the above procedure does not
work.  The first
diagram in Fig.~\ref{fig1}(b) yields $\Sigma=p$.  Inserting this result for
$\Sigma$ into the next two diagrams leads to $\Sigma=p-p^{1/2}$, so that
the expansion in $V$ does not lead to an expansion in $p$.  To correct
this, at ``leading order'' we instead sum up all diagrams involving a {\em single}
link.  The first terms in this sum are shown in the first pair of
parenthesis in Fig.~\ref{fig1}(b).  The
infinite sum yields $\Sigma=p (1-2 [G]_{ii} + 4 [G]_{ii}^2 - ...)=
p/(1+2 [G]_{ii})$.  Solving this equation self-consistently to lowest order
in $p$ yields
$\Sigma=p^2$.  Physically, since the density of links is small, of order $p$,
this result consists of exactly solving the interaction with a single link.

Having done this infinite summation, we
can now consider another infinite series of diagrams (containing two links),
starting with the last one shown in Fig.~\ref{fig1}(b), and
adding additional diagrams where each
single interaction with the given link is replaced by two, three, or more
interactions with the link, as well as diagrams in which crosses are
replaced by circles.  Let us determine the order of this summation in $p$.
The infinite summation replaces the interaction
with a given link with the sum: $1-2 [G]_{ii}+4 [G]_{ii}^2-...
=1/(1+2 [G]_{ii})\propto p$.  Thus, the diagram has four such ``scatterings'', 
leading to a result of order $p^4$. There is an additional factor of $p^2$, 
due to the appearance of two sets of circles/crosses connected by dashed lines.
There is a further factor of $[G]_{ii}^3\propto p^{-3}$ from the three
Green's functions.  Finally, there is a summation over the spacing between
the two impurities; this spacing is of order $p^{-1}$.  As a result,
this sum yields a result which is again of order
$p^2$.  Note that these diagrams lead to a self-energy $\Sigma_{ij}$ which
is off-diagonal in real-space, and hence is momentum dependent; we
consider here the zero momentum part of $\Sigma$.
More complicated diagrams continue to yield results of order
$p^2$.  As a result, after this resummation, we are able to
determine only that $\Sigma_{\rm soft}$ scales as $p^2$, but not
the exact coefficient, and not the higher order corrections:
\begin{equation}
\Sigma_{\rm soft} \propto p^2
\label{sigma_soft}
\end{equation}

\begin{figure}[t]
\vspace*{2.5truecm}
       \includegraphics{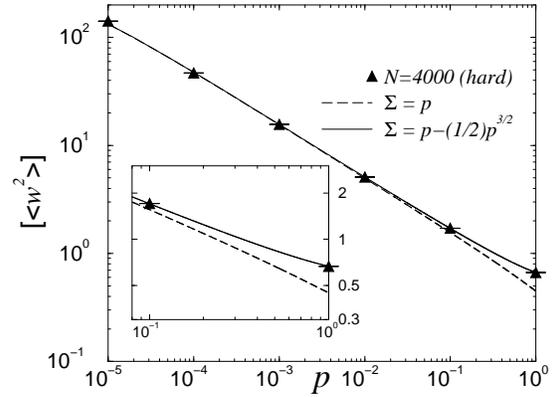}
\vspace*{3.0truecm}
\caption{Comparison of the numerically obtained width (symbols) for the
hard version of the small-world network for
$N$$=$$4000$ with the lowest (dashed lines) and next-to-lowest order (solid
lines) perturbative results. The inset shows a magnified view of the
improvement of the latter.}
\label{fig3}
\end{figure}

Then, the disorder-averaged width can be found from $\Sigma$ by
$\left[\langle w^2 \rangle \right] = \left[ G \right]_{ii} \simeq
\frac{1}{2\sqrt{\Sigma}}$. Thus, the {\em asymptotic} small-$p$
behavior of the width in the thermodynamic limit is $[\langle w^2
\rangle]_{\rm hard} \simeq 1/(2\sqrt{p})$ and $[\langle w^2
\rangle]_{\rm soft} \propto 1/p$, for the respective versions of
the small-world networks. These asymptotic small-$p$, infinite
system-size behaviors are indicated with the two slopes next to
the numerical data in Fig.~\ref{fig2}. In an attempt to match the
behavior of the width for finite systems in the hard
network, we used the
finite-system version of the propagator (see, e.g., \cite{TORO})
with the effective mass from Eq.~(\ref{sigma_hard}).
Note that we ignored various finite-size corrections to $\Sigma$
itself, as being too complicated to be worth calculating. The
results lead to good agreement and are systematically improved by including
the next-to-leading order corrections for the hard network (Fig.~\ref{fig3}).

These results indicate that in both SW versions, the width
approaches a finite value for any non-zero value of $p$ as
$N$$\rightarrow$$\infty$. In the hard network, the scaling of the
width asymptotically approaches that of the mean-field version of
Eq.~(\ref{EW_sw}) \cite{meaf}, while the soft network does not
exhibit this scaling, as expected from the general
criterion \cite{HASTINGS03}. Similarly, for $d$$=$$2$, one finds
$\Sigma_{\rm hard}=p-\ldots$, while for the soft case the
resummation of the expansion enables one to find a logarithmic correction
to the trivial mean-field behavior \cite{future}:
\begin{equation}
\Sigma_{\rm soft}= 2\pi
p/|\log(p)|-\ldots.
\end{equation}
For $d$$>$$2$, $\Sigma$ is asymptotically of
order $p$ for both soft and hard cases, i.e,
the systems are effectively mean-field.

We now interpret the effective mass obtained
perturbatively, Eqs.~(\ref{sigma_hard}) and (\ref{sigma_soft}),
in light of earlier results. The density of
states $\rho(\lambda)$ (the eigenvalue spectrum of the the
coupling matrix $\Gamma$) has been recently investigated in the
context of diffusion on the ``soft'' version of the small-world
network \cite{MONA}. The soft construction of the network allows
for the existence of arbitrarily long ``pure'' chain segments when
the system size $N$ goes to infinity \cite{MONA}. Although the
probability of these quasilinear chain segments of length $l$ is
exponentially small ($\sim e^{-pl}$), they can contribute to
eigenvalues of order $1/l^2$ \cite{LIFS,BRAY}.
Summing up over large $l$ values with the exponential weight above yields
$\rho(\lambda)\sim (1/\sqrt{\lambda})e^{-cp/\sqrt{\lambda}}$
for small $\lambda$ \cite{MONA}, where $c$ is a constant. Thus,
there is no gap in the spectrum. The density of states, however,
vanishes exponentially fast for small $\lambda$ as a result of the
essential singularity in the exponent, and one can refer to $p^2$
as a pseudo-gap. From the right side of
Eq.~(\ref{w2}) it follows that in the $N$$\rightarrow$$\infty$
limit, the disorder-averaged width can be expressed in terms of
the density of states as $[\langle w^2 \rangle ]= \int\!
(1/\lambda)\rho(\lambda)d\lambda$. The small-$\lambda$ behavior of
$\rho(\lambda)$ determines whether the width remains finite or
diverges in the thermodynamic limit.
The exponentially small $\rho(\lambda)$  above
more than compensates for the term $\lambda$ in the
denominator yielding $[\langle w^2 \rangle ]_{\rm soft}\propto 1/p$.

Our perturbative scheme can be extended to calculate
$\rho(\lambda)$ by defining $[G(\lambda)]=P[(\Gamma-\lambda+i\epsilon)^{-1}]$ so
that $\rho(\lambda)=-{\rm Im}[G(\lambda)]/\pi$.
For the hard version of the network, we expect that there is indeed
a true gap of order $p$.  The self-energy is only weakly $\lambda$-dependent.
Each site has one link and for any fixed
segment of the chain, as $N$$\rightarrow$$\infty$, there is a vanishing
probability that any of these links connect two sites in the given segment,
thus preventing the construction of quasilinear segments as above.


We thank Z. Toroczkai, H. Guclu, Z. R\'acz, M.A. Novotny, P.A.
Rikvold, and G. Gy\"orgyi for discussions. G.K. thanks CNLS, Los
Alamos National Laboratory for their hospitality, where this
manuscript was completed. We acknowledge the support of NSF
through Grant No. DMR-0113049 and the support of the Research
Corporation Grant No. RI0761.  MBH was supported by US DOE
W-7405-ENG-36.


\end{document}